\def\year{2022}\relax
\documentclass[letterpaper]{article} % DO NOT CHANGE THIS
\usepackage{aaai22}  % DO NOT CHANGE THIS
\usepackage{times}  % DO NOT CHANGE THIS
\usepackage{helvet}  % DO NOT CHANGE THIS
\usepackage{courier}  % DO NOT CHANGE THIS
\usepackage[hyphens]{url}  % DO NOT CHANGE THIS
\usepackage{graphicx} % DO NOT CHANGE THIS
\usepackage{natbib}  % DO NOT CHANGE THIS AND DO NOT ADD ANY OPTIONS TO IT
\usepackage{caption} % DO NOT CHANGE THIS AND DO NOT ADD ANY OPTIONS TO IT
\DeclareCaptionStyle{ruled}{labelfont=normalfont,labelsep=colon,strut=off} % DO NOT CHANGE THIS
\usepackage{algorithm}
\usepackage{algorithmic}
\usepackage{amsfonts}
\usepackage{amsmath}
\usepackage{longtable}
\usepackage{multirow}
\usepackage{booktabs} 
\usepackage{newfloat}
\usepackage{listings}
\floatstyle{ruled}
\newfloat{listing}{tb}{lst}{}
\floatname{listing}{Listing}
\title{Online Enhanced Semantic Hashing: Towards Effective and Efficient Retrieval \\ for Streaming Multi-Modal Data}
\author {
     Xiao-Ming Wu, Xin Luo\footnote{\noindent Corresponding author.}, Yu-Wei Zhan, Chen-Lu Ding, Zhen-Duo Chen, Xin-Shun Xu
}
\affiliations {
     School of Software, Shandong University\\
   \{wuxiaoming.alg, luoxin.lxin, zhanyuweilif, dingchenlu200103, chenzd.sdu\}@gmail.com, xuxinshun@sdu.edu.cn
}

\usepackage{bibentry}
\begin{document}

\maketitle

\begin{abstract}
With the vigorous development of multimedia equipments and applications, efficient retrieval of large-scale multi-modal data has become a trendy research topic.  Thereinto, hashing has become a prevalent choice due to its retrieval efficiency and low storage cost. Although multi-modal hashing has drawn lots of attention in recent years, there still remain some problems. The first point is that existing methods are mainly designed in batch mode and not able to efficiently handle streaming multi-modal data. The second point is that all existing online multi-modal hashing methods fail to effectively handle unseen new classes which come continuously with streaming data chunks. In this paper, we propose a new model, termed Online enhAnced SemantIc haShing (OASIS). We design novel semantic-enhanced representation for data, which could help handle the new coming classes, and thereby construct the enhanced semantic objective function. An efficient and effective discrete online optimization algorithm is further proposed for OASIS. Extensive experiments show that our method can exceed the state-of-the-art models. For good reproducibility and benefiting the community, our code and data are already publicly available.
\end{abstract}

\section{Introduction}
\noindent With the rapid development of electronic devices and real-world applications, the past decades have witnessed a dramatic increase in the number of multimedia data  \cite{kang2016column,li2016weakly,xu2020federated,li2018deep,zhu2020flexible}. Therefore, efficient retrieval of multimedia data has become a hot research topic. Learning to hash has arisen to be a promising choice because of its fast retrieval speed and low storage consumption \cite{li2020weakly,chen2021long,weng2021online,liu2019supervised}. Roughly speaking, we could divide existing methods into uni-modal hashing \cite{9712384,wang2018supervised,liu2019supervised2,wang2019deep,luo2019discrete}, cross-modal hashing \cite{liu2019ranking,xie2020multi,jin2020deep,nie2020fast,hu2021video}, and multi-modal hashing \cite{liu2012compact,shen2015multi,zhu2020deep}. Thereinto, multi-modal hashing requires that both database and query samples provide heterogeneous multi-modal features. In this paper, we focus on multi-modal hashing which draws a significant need in real-world multimedia retrieval tasks while is relatively unexplored.

\begin{figure*}[t]
\centering
\includegraphics[width=0.95\textwidth]{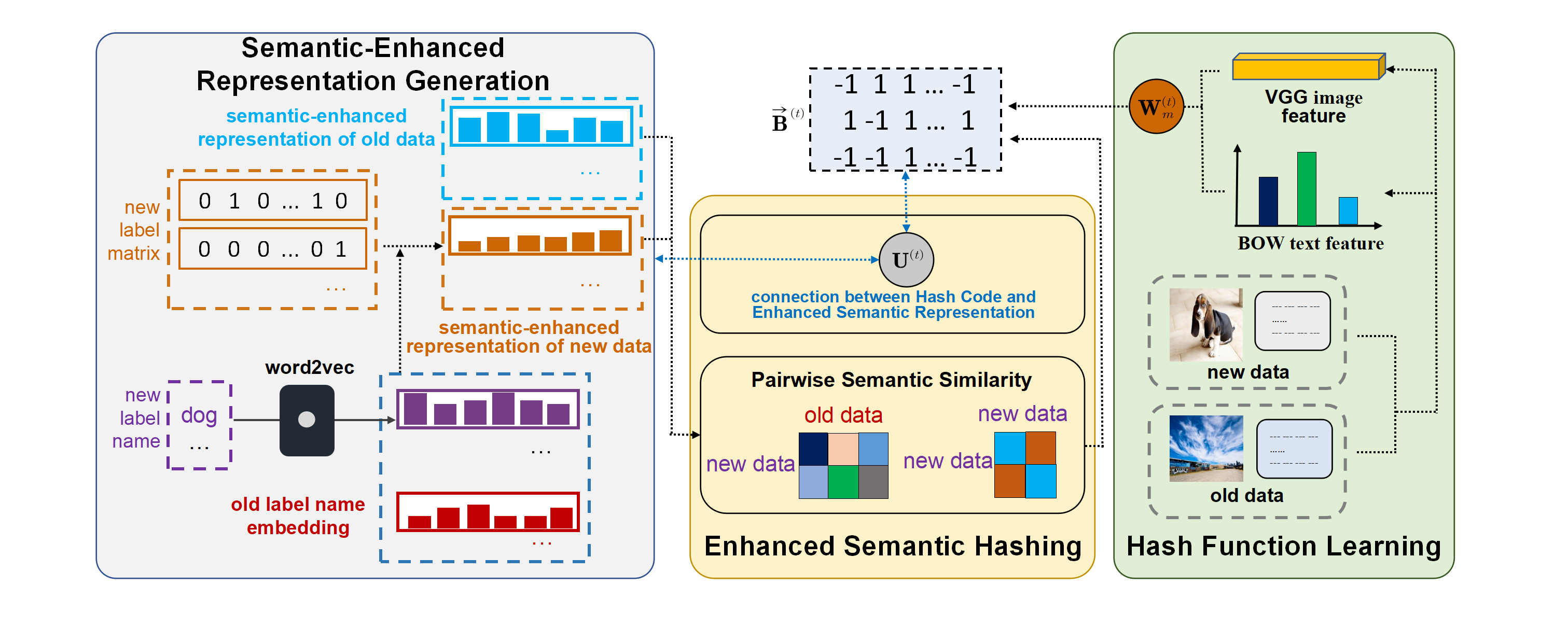}
\caption{The framework of OASIS at the $t$-round. }
\label{framework}
\vspace{-2mm}
\end{figure*}

Despite the promising performance of hashing methods, failing to efficiently learn from streaming data may be an obstacle to applying them to real-world applications. Those batch-based hashing methods assume all training data are available in advance for training and such batch mode needs large space cost. To overcome the limitation, many efforts have been devoted to online hashing. Similarly, we could roughly classify the literature into online uni-modal hashing \cite{huang2013online,cakir2015adaptive,cakir2017mihash,chen2017frosh,weng2020online,tian2019concept,chen2019making}, online cross-modal hashing \cite{xie2016online,qi2017online,wang2020label,yi2021efficient,zhan2022discrete}, and online multi-modal hashing \cite{xie2017dynamic,lu2019flexible}. 

In online hashing literature, one of the most important problems is the class incremental problem, which means new (unknown) categories may appear along with new data chunks. However, most existing works fail to solve it and only a few solutions have been proposed. Some works rise to the challenge by means of offline coding strategies, such as Error Correcting Output Codes \cite{cakir2017online} and Hadamard matrix \cite{lin2018supervised,lin2020hadamard}. Some methods try to increase the number of hash bits for better representation \cite{mandal2018growbit}. Some efforts are made through training multiple complementary hash tables incrementally \cite{tian2020complementary}. Some methods design an end-to-end model to figure out this problem \cite{wu2019deep,chen2019extensible}. However, these solutions are designed for uni-modal and cross-modal hashing while the class incremental problem in online multi-modal hashing is still an open problem without investigations. To the best of our knowledge, the only two online multi-modal hashing methods are Online Dynamic Multi-View Hashing (ODMVH) \cite{xie2017dynamic} and Flexible Online Multi-modal Hashing (FOMH) \cite{lu2019flexible}. However, ODMVH is an unsupervised method and FOMH implicitly assumes that no new categories come with streaming data. In other words, no existing online multi-modal hashing can handle the scenario where new (unknown) categories continually appear along with new data chunks. Besides, we argue that the problem settings of some existing online hashing, which could tackle the class incremental problem, may be impractical: 1) some may relearn the hash codes of old data; 2) some may reuse the original features of old data. When facing large-scale applications, those settings will become inefficient. Hence, in this paper, we formalize the problem by giving some constraints. 1) Multimedia data come in a streaming fashion. 2) The hash codes keep unchanged once learnt and the hash code length is fixed. 3) When updating the hash function, features of new data can be utilized but features of old data chunks are unusable. 4) New (unknown) categories may continually appear along with new data chunks.

To overcome the aforementioned limitations and satisfy the requirements, in this paper, we propose a novel method termed Online enhAnced SemantIc haShing (OASIS), which could learn hash codes and functions when the multimedia instances arrive in streaming fashion. As far as we know, this is the first attempt to investigate the class incremental problem in the context of multi-modal hashing. The main contributions are as follows.
\begin{itemize}
\item \textbf{Task contribution.} We thoroughly investigate the online multi-modal hashing and define a problem setting to better organize this research topic. The proposed task is more challenging and more practical.
\item \textbf{Technical contribution.} This paper conceives a new online multi-modal hashing method and proffers an efficient and effective discrete online optimization algorithm. We generate a novel semantic-enhanced representation for data and thereby construct a semantically enhanced objective function. Besides, with the designed representation, we could handle new categories well. Extensive experiments on two benchmark datasets show that our method can exceed the state-of-the-art models.
\item \textbf{Community contribution.} Our codes are already publicly available\footnote{https://github.com/DravenALG/OASIS}. We hope our work may become one of the enablers for the valuable but relatively unexplored topic and the learning to hash community.
\end{itemize}

% 1) \textbf{Task contribution.} We thoroughly investigate the online multi-modal hashing and define a problem setting to better organize this research topic. The proposed task is more challenging and more practical. 2) \textbf{Technical contribution.} This paper conceives a new online multi-modal hashing method and proffers an efficient and effective discrete online optimization algorithm. We generate a novel semantic-enhanced representation for data and thereby construct a semantically enhanced objective function. Besides, with the designed representation, we could handle new categories well. Extensive experiments on two benchmark datasets show that our method can exceed the state-of-the-art models. 3)  \textbf{Community Contribution.} Our model, training and evaluation code are already available in supplementary material and will be made publicly available. We hope our work may become one of the enablers for the valuable but relatively unexplored topic and the learning to hash community.

\section{Method }
As shown in Figure \ref{framework}, our proposed OASIS contains several modules, i.e., semantic-enhanced representation generation, enhanced semantic hashing, and hash function learning. Details of our OASIS are introduced in this section.

\subsection{Notations and Problem Definition}
In this paper, we assume that multi-modal data comes at a streaming manner. At the $t$-th round, suppose that we have $N^{(t)}$ old multi-modal samples $\widetilde{\mathbf{X}}^{( t )}$, which are observed before current round, and $n^{(t)}$ new ones $\vec{\mathbf{X}}^{(t)}$, where $N^{( t )}=n^{( 1)}+ \cdots + n^{( t-1 )}$. More specifically, as all samples contain features from $M$ modalities, we denote the feature matrix of old data from the $m$-th modality as $\widetilde{\mathbf{X}}_{m}^{( t )}\in R^{d_m\times N^{( t )}}$ and the new samples' feature matrix of the $m$-th modality as $\vec{\mathbf{X}}_{m}^{(t)}\in R^{d_m\times n^{( t )}}$, where $m \in \{ 1, {...}, M \}$ and  $d_m$ is the dimensionality. Meanwhile, we can acquire the labels  $\vec{\mathbf{L}}^{( t )}\in \{ 0,1 \} ^{( c_{n}^{( t )}+c_{o}^{( t )} ) \times n^{( t )}}$ of $\vec{\mathbf{X}}^{( t )}$, where $c_{o}^{(t)}$ is the number of old classes observed in former rounds, $c_{n}^{(t)}$ is the number of new classes which first appear in the $t$-th round and $c_{o}^{(t)}=c_{n}^{(1)}+\cdots +c_{n}^{(t-1)}$. Then, we learn the hash codes $\vec{\mathbf{B}}^{(t)}\in \{ -1,1 \} ^{r\times n^{(t)}}$ for  $\vec{\mathbf{X}}^{(t)}$, where $r$ is the code length. Similarly, $\widetilde{\mathbf{L}}^{( t )}$ and $\widetilde{\mathbf{B}}^{(t)} \in \{ -1,1\} ^{r\times N^{(t)}}=[ \vec{\mathbf{B}}^{(1)},\cdots ,\vec{\mathbf{B}}^{(t-1)}]$  are the label matrix and hash codes of $\widetilde{\mathbf{X}}_{m}^{( t )}$, respectively. 

We let $\mathbf{I}$, $\mathbf{1}$, and $\mathbf{0}$ denote the identity matrix, an all-one matrix, and an all-zero matrix, respectively. Regarding the definition of the operator, we use $\mathrm{tr}( \cdot ) $ to represent the trace operation of the matrix, $\| \cdot \| _F$ to represent the Frobenius form of the matrix, and $\mathrm{sign}( \cdot ) $ to represent the sign function.

In this paper, we focus on the crucial but understudied task, i.e., online multi-modal hashing. We hope our model satisfies the following requirements. 1) Multi-modal data come in a streaming fashion ($\vec{\mathbf{X}}^{(1)}$, $\vec{\mathbf{X}}^{(2)}$, $\cdots$ and so on). 2) The hash codes keep unchanged once learnt and the hash code length $r$ is fixed. 3) When updating the hash function, features of new data $\vec{\mathbf{X}}^{(t)}$ can be utilized but  $\widetilde{\mathbf{X}}^{( t )}$ is unusable. 4) New (unknown) categories may continually appear along with new data chunks ($c_{n}^{(t)}\geq 0$ ).

\subsection{Formulation}
\subsubsection{Basic Hashing Formulation.} We start with one of the commonest formulations in hashing \cite{kang2016column}, i.e., $ \| r \mathbf{S}- \mathbf{B}^T \mathbf{B} \| _{F}^{2}$. One of the greatest weaknesses of this formulation comes from the size of the pairwise similarity matrix $\mathbf{S}$ which is square of the number of training samples. To provide a remedy to this dilemma, many efforts have been made and the frequently-used strategy in online hashing literature is redefining $\mathbf{S}$ as,
\begin{equation}\label{S-orginal}
\mathbf{S}=2\mathbf{P}^{ T}\mathbf{P}-\mathbf{1}\mathbf{1}^T,
\end{equation}
where $\mathbf{P}$ is the 2-norm normalized label matrix defined as $\mathbf{P}_{j}=\mathbf{L}_{j}/ \| \mathbf{L}_{j} \|$, $\| \cdot \|$ represents the length of the vector, $\mathbf{L}_{j}$ and $\mathbf{P}_{j}$ represents the j-th column of $\mathbf{L}$ and $\mathbf{P}$, respectively. Then, the computational complexity can be reduced from square to linear. For the sake of accommodating to online streaming data, pairwise  $\mathbf{S}^{( t )}$, whose size is $(N^{(t)}+n^{(t)})\times(N^{(t)}+n^{(t)})$, at the $t$-th round could be split into four parts, 
\begin{equation} \label{S-spilt}
\begin{aligned}
\begin{array}{l}
\mathbf{S}_{oo}^{\left( t \right)}=2\widetilde{\mathbf{P}}^{\left( t \right) T}\widetilde{\mathbf{P}}^{\left( t \right)}-\mathbf{1}\mathbf{1}^T,
 \mathbf{S}_{no}^{\left( t \right)}=2\vec{\mathbf{P}}^{\left( t \right) T}\widetilde{\mathbf{P}}^{\left( t \right)}-\mathbf{1}\mathbf{1}^T\\
	\mathbf{S}_{on}^{\left( t \right)}=2\widetilde{\mathbf{P}}^{\left( t \right) T}\vec{\mathbf{P}}^{\left( t \right)}-\mathbf{1}\mathbf{1}^T, 
	\mathbf{S}_{nn}^{\left( t \right)}=2\vec{\mathbf{P}}^{\left( t \right) T}\vec{\mathbf{P}}^{\left( t \right)}-\mathbf{1}\mathbf{1}^T,
\end{array}
\end{aligned}
\end{equation}
where $\mathbf{S}_{oo}^{(t)}$ is the semantic similarity between old data, $\mathbf{S}_{on}^{(t)}$ is the semantic similarity between old data and new data, $\mathbf{S}_{no}^{(t)}$ is the semantic similarity between new data and old data, $\mathbf{S}_{nn}^{(t)}$ is the semantic similarity between new data, $\vec{\mathbf{P}}^{\left( t \right)}$ is the 2-norm normalized label matrix of new data and $\widetilde{\mathbf{P}}^{\left( t \right)}$ is the 2-norm normalized label matrix of old data. The only supervised online multi-modal hashing \cite{lu2019flexible} keeps $\mathbf{S}_{nn}^{\left( t \right)}$ only and omits other three. Such strategy may lose the information of old data. The state-of-the-art online cross-modal hashing \cite{wang2020label} learns from the above four parts and is endowed with impressive results. However, it is impossible to handle class incremental problem because the incorrect dimensions for matrix multiplication error will happen when performing $\vec{\mathbf{P}}^{\left( t \right) T}\widetilde{\mathbf{P}}^{\left( t \right)}$ operation if new classes come. In other words, $\mathbf{S}_{on}^{\left( t \right)}$ and $\mathbf{S}_{no}^{\left( t \right)}$ cannot be calculated due to the different number of the classes of the new and old data.

We propose a novel solution to tackle the class incremental problem which is detailedly shown below.

\subsubsection{Semantic-Enhanced Representation.} 
Label names are often naturally well separated from each other and contain good expressions of category-specific semantics \cite{wang2018joint}. Thus, we introduce the embedding of label names into our framework to gain more powerful semantic information to guide the hash learning. 

In online setting, as new labels may continually appear along with new data chunk, we use  $\mathbf{K}_{o}^{(t)} \in \mathbb{R}^{ c_o^{(t)} \times f} $ to represent the old  labels' name embeddings and ${\mathbf{K}}_{n}^{(t)}\in  \mathbb{R}^{ c_n^{(t)} \times f}$ to denote new labels' name embeddings, where $f$ is the dimensionality of word2vec vector. By concatenating them, the overall label name embeddings can be obtained ${\mathbf{K}}^{(t)} \in \mathbb{R}^{( c_{n}^{(t)}+c_o^{(t)} ) \times f}$ . In our work, Google's word2vec model \cite{DBLP:journals/corr/abs-1301-3781} is leveraged and $f=300$.

Then, we propose a semantic-enhanced matrix  $ \vec{\mathbf{G}}^{(t)}\in R^{f\times n^{(t)}}$ that comprises both label matrix and label name embeddings. The semantic-enhanced matrix is defined as,
\begin{equation}\label{semantic_enhanced_label_matrix}
 \vec{\mathbf{G}}_{j}^{(t)}=\frac{{\mathbf{K}}^{(t) T}\vec{\mathbf{L}}_{j}^{(t)}}{\| {\mathbf{K}}^{(t) T}\vec{\mathbf{L}}_{j}^{(t)} \|}, j=1,2\cdots n^{(t)},
\end{equation}
where $\vec{\mathbf{L}}_{j}^{(t)}$ represents the $j$-th column of $\vec{\mathbf{L}}^{(t)}$, $ \vec{\mathbf{G}}_{j}^{(t)}$ represents the $j$-th column of $ \vec{\mathbf{G}}^{(t)}$, and $\| \cdot \|$ represents the length of vectors. Furthermore, we turn each column of the matrix into a unit vector, which can easily express the cosine similarity when calculating $\mathbf{S}^{(t)}$. The semantic-enhanced matrix offers instances semantic-rich representations and  plays an important role in guiding the learning of OASIS.

\subsubsection{Enhanced Semantic Hashing.} 
Now, let's revisit Eq.(\ref{S-spilt}). With the semantic-enhanced representation of instances, we could redefine the pairwise semantic similarity as follows, 
\begin{equation}\label{S-split-new}
\begin{aligned}
\begin{array}{l}
\mathbf{S}_{oo}^{(t)}= \widetilde{\mathbf{G}}^{(t) T}\widetilde{\mathbf{G}}^{(t)},	\mathbf{S}_{no}^{(t)}= \vec{\mathbf{G}}^{(t) T}\widetilde{\mathbf{G}}^{(t)},\\
\mathbf{S}_{on}^{(t)}=\widetilde{\mathbf{G}}^{(t) T} \vec{\mathbf{G}}^{(t)}, \mathbf{S}_{nn}^{(t)}=\vec{\mathbf{G}}^{(t) T} \vec{\mathbf{G}}^{(t)},\\
\end{array}
\end{aligned}
\end{equation}
where $\widetilde{\mathbf{G}}^{(t)}\in \mathbb{R}^{f\times N^{(t)}}$ denotes the semantic-enhanced representation of old samples and can be obtained before round $t$. It is worth noting that Eq.(\ref{S-split-new}) could naturally handle the class incremental problem. As the size of $\vec{\mathbf{G}}^{(t)}$ is $f\times n^{(t)}$ and the the size of $\widetilde{\mathbf{G}}^{(t)}$ is $f\times N^{(t)}$, the dimensions for matrix multiplication is correct. We elaborately bypass the obstacle mentioned above and could construct the connections between new and old classes. Then, we propose the loss guided by enhanced semantic pairwise similarity as,
\begin{equation}
\begin{aligned}
&\min_{ \vec{\mathbf{B}}^{(t)}\in \{ -1,1\} ^{r\times n^{(t)}} }  	\| r\mathbf{S}_{nn}^{(t)}- \vec{\mathbf{B}}^{(t) T}\vec{\mathbf{B}}^{(t)} \| _{F}^{2} \\&  +\| r\mathbf{S}_{on}^{(t)}-\widetilde{\mathbf{B}}^{(t) T}\vec{\mathbf{B}}^{(t)} \| _{F}^{2}+\| r\mathbf{S}_{no}^{(t)}- \vec{\mathbf{B}}^{(t) T}\widetilde{\mathbf{B}}^{(t)} \| _{F}^{2},
\end{aligned}
\end{equation}
where $\mathbf{S}_{oo}^{(t)}$ is omitted because to-be-learnt $ \vec{\mathbf{B}}^{(t)}$ is irrelevant to it. In this equation, knowledge of both old and new classes can be embedded and may constructively guide the hash learning.

To further enhance the semantic information that guides the hash learning, we also directly construct the connection between hash codes and semantic-enhanced representation,
\begin{equation}\label{G-B}
\begin{aligned}
&\min_{ \vec{\mathbf{B}}^{(t)},\mathbf{U}^{(t)} } \|  \vec{\mathbf{G}}^{(t)}-\mathbf{U}^{(t)}\vec{\mathbf{B}}^{(t)} \| _{F}^{2}+\| \widetilde{\mathbf{G}}^{(t)}-\mathbf{U}^{(t)}\widetilde{\mathbf{B}}^{(t)} \| _{F}^{2}
\\& + \theta \| \mathbf{U}^{(t)} \| _{F}^{2}, \ \ s.t.\vec{\mathbf{B}}^{(t)}\in \{ -1,1\} ^{r\times n^{(t)}},
\end{aligned}
\end{equation}
where $\mathbf{U}^{(t)}$ is a mapping matrix and $\theta$ is a hyperparameter controlling the regularization term. Here, we use the same mapping matrix for both old and new data in order that knowledge previously learnt could be compatible with new knowledge. Note that, although this loss looks like the widely-used term which uses hash codes for classification, the idea behind Eq.(\ref{G-B}) is totally different.
%and we try to alleviate the concept drift problem

\subsubsection{Hash Function Learning.}
Currently, linear hash functions occupy the mainstream position in online hashing domain while neural network based methods are extremely scarce. The possible reasons may be that: 1) Online hashing focuses more on efficiency as almost all publications report the training time comparison results; 2) Training of neural network based hash functions is much more time-consuming than the training of linear functions. Following the vast majority, we design our hash function learning module using the efficient and straightforward linear mapping as follows,
\begin{equation}\label{function}
\begin{aligned}
&\underset{ \vec{\mathbf{B}}^{(t)}\in \{ -1,1\} ^{r\times n^{(t)}},\mathbf{W}_m^{(t)}}{\min} \sum_{m=1}^M( \|  \vec{\mathbf{B}}^{(t)} -\mathbf{W}_m^{(t)}   \vec{\mathbf{X}}_{m}^{(t)} \| _{F}^{2}\\& +\| \widetilde{\mathbf{B}}^{(t)}-\mathbf{W}_m^{(t)}\widetilde{\mathbf{X}}_{m}^{(t)} \| _{F}^{2}    +\delta \| \mathbf{W}_m^{(t)} \| _{F}^{2} ), 
\end{aligned}
\end{equation}
where $M$ represents the total number of modalities, $\mathbf{W}_m^{(t)}$ is the projection of the $m$-th modality, and $\delta$ is a hyperparameter balancing the regularization term. Although Eq.(\ref{function}) is simple, it has several advantages beyond simplicity and efficiency: 1)This equation reduces the quantization loss between the learnt hash code and the real-valued mapping results; 2) The catastrophic forgetting problem could be alleviated through simultaneously considering old and new data. Although above loss contains $\widetilde{\mathbf{X}}_{m}^{(t)}$, our method still meets the third requirement as can be seen in \textbf{Algorithm \ref{alg:algorithm}} whose inputs do not need features of old data chunks. This is because of the proposed novel online optimization which is presented below.

\subsubsection{Overall Objective Function.}
In summary, by combining all metioned modules together and creating reasonable modifications, the total loss function of OASIS is shown below,
\begin{equation}\label{overall-loss}
\begin{aligned}
&\min_{\vec{\mathbf{B}}^{(t)}\in \{ -1,1\} ^{r\times n^{(t)}},\vec{\mathbf{V}}^{(t)},\mathbf{U}^{(t)},\mathbf{W}_m^{(t)}} \|  \vec{\mathbf{B}}^{(t)}-\vec{\mathbf{V}}^{(t)} \| _{F}^{2}+ 
\\ &\alpha(  \| r\mathbf{S}_{nn}^{(t)}- \vec{\mathbf{B}}^{(t) T}\vec{\mathbf{V}}^{(t)} \| _{F}^{2}   +\| r\mathbf{S}_{on}^{(t)}-\widetilde{\mathbf{B}}^{(t) T}\vec{\mathbf{V}}^{(t)} \| _{F}^{2}
\\ & +\| r\mathbf{S}_{no}^{(t)}- \vec{\mathbf{B}}^{(t) T}\widetilde{\mathbf{V}}^{(t)} \| _{F}^{2} )  +\beta ( \|  \vec{\mathbf{G}}^{(t)}-\mathbf{U}^{(t)}\vec{\mathbf{V}}^{(t)} \| _{F}^{2}
 \\& +\| \widetilde{\mathbf{G}}^{(t)}-\mathbf{U}^{(t)}\widetilde{\mathbf{V}}^{(t)} \| _{F}^{2}+\theta \| \mathbf{U}^{(t)} \| _{F}^{2} ) +\gamma \sum_{m=1}^M( \|  \vec{\mathbf{B}}^{(t)}
\\ & - \mathbf{W}_m\vec{\mathbf{X}}_{m}^{(t)} \| _{F}^{2}+\| \widetilde{\mathbf{B}}^{(t)}-\mathbf{W}_m^{(t)}\widetilde{\mathbf{X}}_{m}^{(t)} \| _{F}^{2}+ {\delta \|\mathbf{W}_m \| _{F}^{2} )}
\\ & s.t.  ,\vec{\mathbf{V}}^{(t)}\vec{\mathbf{V}}^{(t) T}=n^{(t)}\mathbf{I}_r, \vec{\mathbf{V}}^{(t)}\mathbf{1}^T=\mathbf{0},
\end{aligned}
\end{equation}
where  $\vec{\mathbf{V}}^{(t)}$ is the real-valued approximation of hash code with several constraints, $\alpha$, $\beta$, and $\gamma$ are tradeoff parameters. By using $\vec{\mathbf{V}}^{(t)}$ to approximate $ \vec{\mathbf{B}}^{(t)}$, the hard optimization problem of $\vec{\mathbf{B}}^{(t)}$ can be simplified. Besides,  the constraint of $\vec{\mathbf{V}}^{(t)}\vec{\mathbf{V}}^{(t) T}=n^{(t)}\mathbf{I}_r$ can make each bit represent as much information as possible, and the $\vec{\mathbf{V}}^{(t)}\mathbf{1}^T=\mathbf{0}$ constraint can make the hash code more discriminative.

\subsection{Online Optimization}
We propose a novel discrete online optimization for Eq.(\ref{overall-loss}) to learn hash codes and function at the $t$-th round. Specifically, we optimize the variables one by one and repeat the procedure several times until convergence. At round $t$, the optimization steps are presented in the following.

\textbf{Step 1: Update $\vec{\mathbf{V}}^{(t)}$}. When other variables remain unchanged, the optimization problem of $\vec{\mathbf{V}}^{(t)}$ can be simplified to the following form.
\begin{equation}\label{Z-sub-problem}
\begin{aligned}
&\underset{\vec{\mathbf{V}}^{(t)}}{\max}\,\,\mathrm{tr}(\mathbf{Z}\vec{\mathbf{V}}^{(t) T} ) ,
\\
&s.t. \vec{\mathbf{V}}^{(t)}\vec{\mathbf{V}}^{(t) T}=n^{(t)}\mathbf{I}_r, \vec{\mathbf{V}}^{(t)}\mathbf{1}^T=\mathbf{0},
\end{aligned}
\end{equation}
where $\mathbf{Z}=\alpha r\mathbf{D}_{1}^{(t)}   \vec{\mathbf{G}}^{(t)}+ \vec{\mathbf{B}}^{(t)}+\gamma \mathbf{U}^{(t)T} \vec{\mathbf{G}}^{(t)}$ and $\mathbf{D}_{1}^{(t)}= \vec{\mathbf{B}}^{(t)} \vec{\mathbf{G}}^{(t) T}+\mathbf{D}_{1}^{( t-1 )}$. Notably, $\mathbf{D}_{1}$ is an \textbf{intermediate variable}. By temporarily storing $\mathbf{D}_{1}^{( t-1 )}$, which is calculated at the previous $(t-1)$-th round, and directly using it at the current $t$-th round, the calculation of $\mathbf{D}_{1}^{(t)}$ can be very efficient. %For the first round, $\mathbf{D}_{1}^{(1)}= \vec{\mathbf{B}}^{(1)} \vec{\mathbf{G}}^{(1) T}$. 
%\begin{equation}
%\begin{aligned}
%\mathbf{D}_{1}^{(t)}= \vec{\mathbf{B}}^{(t)} \vec{\mathbf{G}}^{(t) T}+\mathbf{D}_{1}^{( t-1 )}.
%\end{aligned}
%\end{equation}

Let us return to the optimization of $\vec{\mathbf{V}}^{(t)}$. The simplified optimization problem in Eq.(\ref{Z-sub-problem}) is similar to the form in \cite{liu2014discrete} and can be optimized as follows. First, we perform the eigenvalue decomposition of $\mathbf{ZJZ}^T$ and the formula is as follows,
\begin{equation}
\begin{aligned}
\mathbf{ZJZ}^T=[ \mathbf{O}\,\,\overline{\mathbf{O}} ] [ \begin{matrix}
 \mathbf{\Sigma} ^2&  \mathbf{0}\\
 \mathbf{0}&  \mathbf{0}\\
\end{matrix} ] [ \mathbf{O}\,\,\overline{\mathbf{O}} ] ,
\end{aligned}
\end{equation}
where $\mathbf{J}=\mathbf{I}-\frac{1}{n^{\left( t \right)}}\mathbf{1}\mathbf{1}^T$. Then we calculate $\mathbf{N}=\mathbf{JZ}^T\mathbf{O}\mathbf{\Sigma} ^{-1}\in R^{n^{\left( t \right)}\times r'}$, where $r'$ is the number of positive eigenvalues. The $\overline{\mathbf{N}}\in R^{n^{\left( t \right)}\times ( r-r' )}$ is initially set to a random matrix followed by the Gram-Schmidt orthogonalization. Finally, we can get the solution of $\vec{\mathbf{V}}^{(t)}$,
\begin{equation}\label{V}
\begin{aligned}
\vec{\mathbf{V}}^{(t)}=\sqrt{n^{(t)}}[ \mathbf{O}\,\,\overline{\mathbf{O}} ] [ \mathbf{N}\,\,\overline{\mathbf{N}} ] ^T.
\end{aligned}
\end{equation}

\textbf{Step 2: Update $\mathbf{U}^{(t)}$.} Fixing other variables and setting the derivative of Eq.(\ref{overall-loss}) w.r.t. $\mathbf{U}^{(t)}$ to zero, the update formula can be obtained as follows,
\begin{equation}\label{U}
\begin{aligned}
\mathbf{U}^{(t)}=\mathbf{D}_{2}^{(t)}( \mathbf{D}_{3}^{(t)}+\theta \mathbf{I}) ^{-1},
\end{aligned}
\end{equation}
where $\mathbf{D}_{2}^{(t)}= \vec{\mathbf{G}}^{(t)}\vec{\mathbf{V}}^{(t) T}+\mathbf{D}_{2}^{( t-1 )}$ and $\mathbf{D}_{3}^{(t)}=\vec{\mathbf{V}}^{(t)}\vec{\mathbf{V}}^{(t) T}+\mathbf{D}_{3}^{( t-1 )}$. Notably, $\mathbf{D}_{2}$ and $\mathbf{D}_{3}$ are \textbf{intermediate variables}. By temporarily storing them at last round and directly using them at current round, the optimization can be extremely efficient.
%\begin{equation}
%\begin{aligned}
%\mathbf{D}_{2}^{(t)}= \vec{\mathbf{G}}^{(t)}\vec{\mathbf{V}}^{(t) T}+\mathbf{D}_{2}^{( t-1 )},
%\\
%\mathbf{D}_{3}^{(t)}=\vec{\mathbf{V}}^{(t)}\vec{\mathbf{V}}^{(t) T}+\mathbf{D}_{3}^{( t-1 )}.
%\end{aligned}
%\end{equation}

\textbf{Step 3: Update $ \vec{\mathbf{B}}^{(t)}$.} When other variables are fixed, the generation of hash codes can be simplified as,
\begin{equation}
\begin{aligned}
\underset{ \vec{\mathbf{B}}^{(t)}}{\max}\,\,\mathrm{tr}( \mathbf{Q} \vec{\mathbf{B}}^{(t) T} ), \  \ \ s.t. \ \vec{\mathbf{B}}^{(t)}\in \{ -1,1\} ^{r\times n^{(t)}},
\end{aligned}
\end{equation}
where $\mathbf{Q}=\alpha r\mathbf{D}_{4}^{(t)}   \vec{\mathbf{G}}^{(t)}+\vec{\mathbf{V}}^{(t)}+\gamma \sum_{m=1}^M{\mathbf{W}_m^{(t)}}\vec{\mathbf{X}}_{m}^{(t)}$ and $\mathbf{D}_{4}^{(t)}=\vec{\mathbf{V}}^{(t)} \vec{\mathbf{G}}^{(t) T}+\mathbf{D}_{4}^{( t-1 )}$. Notably, similar with $\mathbf{D}_{1}$, $\mathbf{D}_{2}$, and $\mathbf{D}_{3}$, $\mathbf{D}_{4}$ is called \textbf{intermediate variable} and is helpful for effcient opitmization. 
%\begin{equation}
%\begin{aligned}
%\mathbf{D}_{4}^{(t)}=\vec{\mathbf{V}}^{(t)} \vec{\mathbf{G}}^{(t) T}+\mathbf{D}_{4}^{( t-1 )}.
%\end{aligned}
%\end{equation}

Because of $\mathrm{tr}(\mathbf{Q}\vec{\mathbf{B}}^{(t) T} ) =\sum_{i,j}{\mathbf{Q}_{ij} \vec{\mathbf{B}}_{ij}^{(t)}}$ and $ \vec{\mathbf{B}}^{(t)}\in \{ -1,1\} ^{r\times n^{(t)}}$, we can let $ \vec{\mathbf{B}}_{ij}=1$ when $\mathbf{Q}_{ij}$ is positive and $ \vec{\mathbf{B}}_{ij}=-1$ when $\mathbf{Q}_{ij}$ is negative to maximize the simplified optimization function, where $i$ and $j$ represent the $i$-th row and $j$-th column of the matrix. Therefore, we can get the following formula,
\begin{equation}\label{B}
\begin{aligned}
 \vec{\mathbf{B}}^{(t)}=\mathrm{sign}(\mathbf{Q}).
\end{aligned}
\end{equation}

\textbf{Step 4: Update $\mathbf{W}_m^{(t)}$}. Similar to the update of  $\mathbf{U}^{(t)}$, the update formula of  $\mathbf{W}_m^{(t)}$ can be given as follows,
\begin{equation}\label{W}
\begin{aligned}
\mathbf{W}_m^{(t)}=\mathbf{D}_{5}^{(t)}( \mathbf{D}_{6}^{(t)}+\delta \mathbf{I} ) ^{-1},
\end{aligned}
\end{equation}
where $\mathbf{D}_{5}^{(t)}= \vec{\mathbf{B}}^{(t)}\vec{\mathbf{X}}_{m}^{(t) T}+\mathbf{D}_{5}^{( t-1 )}$ and $\mathbf{D}_{6}^{(t)}=\vec{\mathbf{X}}_{m}^{\left( t \right)}\vec{\mathbf{X}}_{m}^{\left( t \right) T}+\mathbf{D}_{6}^{( t-1 )}$. Here, we could temporarily store $\mathbf{D}_{5}^{( t-1 )}$ and $\mathbf{D}_{6}^{( t-1 )}$ at the last round and   use them at  $t$-th round to get $\mathbf{D}_{5}^{( t )}$ and $\mathbf{D}_{6}^{( t)}$ directly. In light of these intermediate variables, efficiency of the optimization can be ensured.
%\begin{equation}
%\begin{aligned}
%\mathbf{D}_{5}^{(t)}= \vec{\mathbf{B}}^{(t)}\vec{\mathbf{X}}_{m}^{(t) T}+\mathbf{D}_{5}^{( t-1 )},
%\\
%\mathbf{D}_{6}^{(t)}=\overrightarrow{\mathbf{X}}_{m}^{\left( t \right)}\overrightarrow{\mathbf{X}}_{m}^{\left( t \right) T}+\mathbf{D}_{6}^{( t-1 )}.
%\end{aligned}
%\end{equation}

\subsubsection{Overall Algorithm.}
For better understanding, we summarize the proposed online optimization in \textbf{Algorithm \ref{alg:algorithm}}. It's worth noting that although we can find  $\widetilde{\mathbf{X}}_{m}^{(t)}$ in the objective function, the old feature is not needed in the real optimization.  Our model satisfies all four requirements raised in the Problem Definition Section.

Besides, the sizes of $\mathbf{D}_{i}^{(t-1)}$ ($i=\{1,\cdots ,6\})$ are ${r\times f}$, ${f\times r}$, ${r\times r} $, ${r\times f} $, ${r\times d_m} $, and ${d_m\times d_m}$. Although these matrices contain rich information from old data, they are irrelevant with $N^{(t)}$. 

%$\mathbf{D}_{1}^{\left( t \right)}\in \mathbb{R} ^{r\times f}$
%$ \mathbf{D}_{2}^{\left( t \right)}\in \mathbb{R} ^{f\times r}$
%$ \mathbf{D}_{3}^{\left( t \right)}\in \mathbb{R} ^{r\times r} $
%$ \mathbf{D}_{4}^{\left( t \right)}\in \mathbb{R} ^{r\times f} $
%$ \mathbf{D}_{5}^{\left( t \right)}\in \mathbb{R} ^{r\times d_m} $
%$ \mathbf{D}_{6}^{\left( t \right)}\in \mathbb{R} ^{d_m\times d_m}$

\subsubsection{Complexity Analysis.}
The complexity of updating $\vec{\mathbf{V}}^{(t)}$ is $O( ( 10r+3r^2+3rf+f ) n^{( t )}+2rf+r^3 )$ , the complexity of getting $\mathbf{U}^{(t)}$ is $O( ( 2r+fr+f+r^2 ) n^{( t )}+2r^2+r^3+fr^2 ) $, the complexity of learning $\vec{\mathbf{B}}^{(t)}$ is $O( ( 5r+2fr+f+\sum_{m=1}^M{rd_m} ) n^{( t )}+rf ) $, and the complexity of generating $\mathbf{W}_m^{(t)}$ is $O( M( d_{m}^{2}+rd_m+d_m )n^{( t )}+M( 2d_{m}^{2}+d_{m}^{3}+rd_{m}^{2}+rd_m ) ) $. It is worth noting that variables $\mathbf{D}_{i}^{(t-1)}$ ($i=\{1,\cdots ,6\}$) can be obtained at the previous round and used directly at this round. Thus, there is no need to consider the time complexity of calculating them.

From the above analysis, we can find that the online optimization is linearly related to the size of new data chunk  $n^{\left( t \right)}$ and is irrelevant to the old database  $N^{( t )}$. To conclude, our proposed optimization is efficient and scalable to large-scale online retrieval applications.

\begin{algorithm}[tb]
\caption{The optimization of OASIS at the $t$-th round.}
\label{alg:algorithm}
\textbf{Input}: $\vec{\mathbf{X}}_{m}^{(t)}$ ($m={1,2...M}$), $\vec{\mathbf{L}}^{(t)}$, $\vec{\mathbf{K}}^{(t)}$, and $\mathbf{D}_{i}^{(t-1)}$ ($i=\{1,\cdots ,6\}$).\\
\textbf{Output}: $\mathbf{D}_{i}^{(t)}$ ($i=\{1,\cdots ,6\}$), $\mathbf{B}^{(t)}$, and hash function.\\
\textbf{Main Algorithm}:
\begin{algorithmic}[1]
\STATE  Randomly initialize $\mathbf{U}^{( t )}$, $\vec{\mathbf{B}}^{(t)}$, and  $\mathbf{W}_m^{( t )}$.
\WHILE{not converged or not reach the max iterations}
\STATE Update $\vec{\mathbf{V}}^{(t)}$ with Eq.(\ref{V}); save $\mathbf{D}_{1}^{(t)}$.
\STATE Update $\mathbf{U}^{( t )}$ with Eq.(\ref{U}); save $\mathbf{D}_{2}^{(t)}$ and $\mathbf{D}_{3}^{(t)}$.
\STATE Update $\vec{\mathbf{B}}^{(t)}$ with Eq.(\ref{B}); save $\mathbf{D}_{4}^{(t)}$.
\STATE Update $\mathbf{W}_m^{( t )}$ with Eq.(\ref{W});  save $\mathbf{D}_{5}^{(t)}$ and $\mathbf{D}_{6}^{(t)}$.
\ENDWHILE
\end{algorithmic}
\end{algorithm}

\subsection{Retrieval Precedure}
When query sample comes, we first learn its hash code  $\mathbf{b}_{query}=\mathrm{sign}( \sum_{m=1}^M{\mathbf{W}_m^{( t )}\mathbf{x}_{query-m}} ) $, 
where $\mathbf{W}_m^{( t )}$ is the up to date hashing projection and $\mathbf{x}_{query-m}$ represents query's feature of the $m$-th modality.

Then, we can calculate the Hamming distance of hash codes between query and database to measure the similarity. Those instances, which are considered to be similar, can be returned as retrieval results.

\begin{table*}[]
\caption{MAP results at the last round.}
\label{map_last_round}
\begin{center} \footnotesize
\begin{tabular}{ccllllllll}
\hline
\multirow{2}{*}{Settings}           & \multirow{2}{*}{Methods} & \multicolumn{4}{c}{MIRFlickr}                                                                                  & \multicolumn{4}{c}{NUS-WIDE}                                                                                   \\ \cline{3-10} 
                                    &                          & \multicolumn{1}{c}{16bit} & \multicolumn{1}{c}{32bit} & \multicolumn{1}{c}{64bit} & \multicolumn{1}{c}{128bit} & \multicolumn{1}{c}{16bit} & \multicolumn{1}{c}{32bit} & \multicolumn{1}{c}{64bit} & \multicolumn{1}{c}{128bit} \\ \hline
\multirow{4}{*}{The first setting}  & FOMH \cite{lu2019flexible}                     & 0.6007                    & 0.6115                    & 0.6135                    & 0.6127                     & 0.5187                    & 0.5716                    & 0.5670                    & 0.5668                     \\
                                    & OMH-DQ \cite{Lu0CNZ19}                   & 0.6520                    & 0.6628                    & 0.6773                    & 0.6875                     & 0.5084                    & 0.5517                    & 0.5840                    & 0.5829                     \\
                                    & SAPMH \cite{zheng2020adaptive}                    & 0.6857                    & 0.7034                    & 0.7121                    & 0.7187                     & 0.5773                    & 0.5830                    & 0.6059                    & 0.6098                     \\
                                    & OASIS                    & \textbf{0.8558}           & \textbf{0.8628}           & \textbf{0.8703}           & \textbf{0.8734}            & \textbf{0.7742}           & \textbf{0.7847}           & \textbf{0.7951}           & \textbf{0.7999}            \\ \hline
\multirow{4}{*}{The second setting} & FOMH \cite{lu2019flexible}                     & 0.5132                    & 0.5638                    & 0.5808                    & 0.5945                     & 0.4781                    & 0.4736                    & 0.4853                    & 0.4754                     \\
                                    & OMH-DQ \cite{Lu0CNZ19}                   & 0.5948                    & 0.6206                    & 0.6264                    & 0.6386                     & 0.6965                    & 0.6929                    & 0.7224                    & 0.7324                     \\
                                    & SAPMH \cite{zheng2020adaptive}                    & 0.5986                    & 0.6378                    & 0.6410                    & 0.6465                     & 0.7326                    & 0.7415                    & 0.7552                    & 0.7621                     \\
                                    & OASIS                    & \textbf{0.7833}           & \textbf{0.8008}           & \textbf{0.8023}           & \textbf{0.8066}            & \textbf{0.7931}           & \textbf{0.7885}           & \textbf{0.7867}           & \textbf{0.8126}            \\ \hline
\multirow{4}{*}{The third setting}  & FOMH \cite{lu2019flexible}                     & 0.0598                    & 0.0580                    & 0.0593                    & 0.0706                     & 0.0118                    & 0.0104                    & 0.0139                    & 0.0159                     \\
                                    & OMH-DQ \cite{Lu0CNZ19}                   & 0.1068                    & 0.1283                    & 0.1385                    & 0.1656                     & 0.0134                    & 0.0166                    & 0.0209                    & 0.0248                     \\
                                    & SAPMH \cite{zheng2020adaptive}                    & 0.0781                    & 0.1064                    & 0.1123                    & 0.1142                     & 0.0120                    & 0.0145                    & 0.0175                    & 0.0202                     \\
                                    & OASIS                    & \textbf{0.1595}           & \textbf{0.2838}           & \textbf{0.2557}           & \textbf{0.2197}            & \textbf{0.1211}           & \textbf{0.2481}           & \textbf{0.2775}           & \textbf{0.2628}            \\ \hline
\end{tabular}
\end{center}
\end{table*}

\section{Experiment}
\subsection{Experimental Settings}
\subsubsection{Datasets.}
In this paper, we chose two widely-used datasets, i.e., MIRFlickr \cite{huiskes2008mir} and NUS-WIDE \cite{chua2009nus}. \textbf{MIRFlickr} has  $25,000$ instances and $24$ categories in total. Following \cite{jiang2019discrete}, we removed instances with rare tags that appear less than $20$ times and finally had $20,015$ instances left. Then, we passed images through pre-trained VGG network to obtain $4096$-d deep features and expressed texts as $1386$-d BOW features. \textbf{NUS-WIDE} has $269,648$ instances and $81$ categories. Following the setting of \cite{lu2019flexible}, only the top $21$ most common categories are used and $195,834$ instances are finally left. Similar with MIRFlickr, we fed images into the VGG network to obtain $4096$-d deep features and represented the text modality as $5018$-d BOW features. Features used in this paper can be obtained from {https://github.com/DravenALG/OASIS}.

\subsubsection{Evaluation Protocols.}
Mean Average Precision (MAP) is adopted as the evaluation criterion and larger value indicates better performance. As stated in above sections, during retrieval phase, we calculated the Hamming distance of hash codes between queries and database to measure the similarity. As both MIRFlickr and NUS-WIDE are multi-label datasets, we considered two samples to be similar if they share at least one common label.

Furthermore, we have designed three types of online experimental settings. The {first} online setting assumes that all categories are known before training. In other words, for $t$ from $2$ to the maximum, $c_{n}^{(t)}$ (the number of new classes which first appear at the $t$-th round) is always equal to 0. On the contrary, the second and third settings simulate the scenario where new (unknown) categories  continually appear along with new data chunks ($c_{n}^{(t)}\textgreater 0$ ). For the {second} online setting, new data carries some old categories which first appear before current round and some new categories. For the {third} setting, all categories of new data are new (unknown).

(1) \emph{The first online setting:} For both datasets, we randomly selected $2,000$ samples to form the test set and left the remaining samples as training set. That is, the size of training set of MIRFlickr is $18,015$ and NUS-WIDE has $193,834$ training points. The first 9 chunks of MIRFlickr have the same size of $2,000$ while the $10$-th chunk includes the remaining $15$ samples. For NUS-WIDE, size of the first $19$ chunks is $10,000$ while the remaining $3,834$ samples constitute the $20$-th chunk. All data chunks are constructed by randomly selecting samples. (2) \emph{The second online setting:} MIRFlickr is divided into 10 rounds and NUS-WIDE is divided into $20$ rounds. Specifically, we assigned some labels for each round, ensuring that labels at the new round include the labels appear at previous rounds and at least one new category appear. Then, we selected eligible data according to the set of labels at each round. It is worth noting that the data at the new round would exclude the data at previous rounds to avoid duplication. Subsequently, the training data and test data of each round are randomly selected from the data of each round at a ratio of $9:1$. (3) \emph{The third online setting:} We stipulated that MIRFlickr has $4$ rounds and NUS-WIDE has $8$ rounds. We allocated some labels for each round. Different from the second setting, we set labels at the new round as totally different from the labels at the last rounds. Then, we selected the data based on the labels at each round. The training data and test data of each round are randomly selected from the data of each round at a ratio of $9: 1$. In all settings, we use all seen training data as the database and test data selected at each round as queries. More details can be found in our code.

\begin{figure}[t]
\centering
\includegraphics[width=0.8\columnwidth]{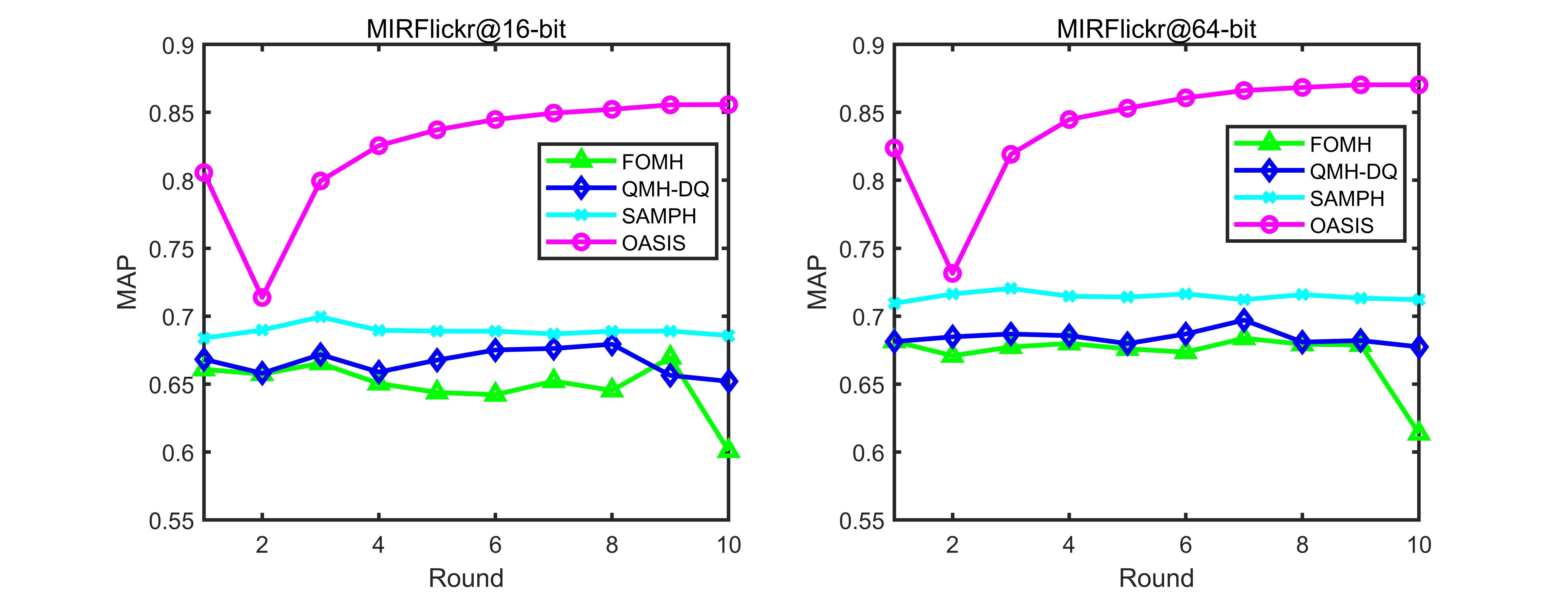}
\caption{ MAP results at all rounds on MIRFlickr. }
\label{MIR_map}
\vspace{-2mm}
\end{figure}

\begin{figure}[t]
\centering
\includegraphics[width=0.8\columnwidth]{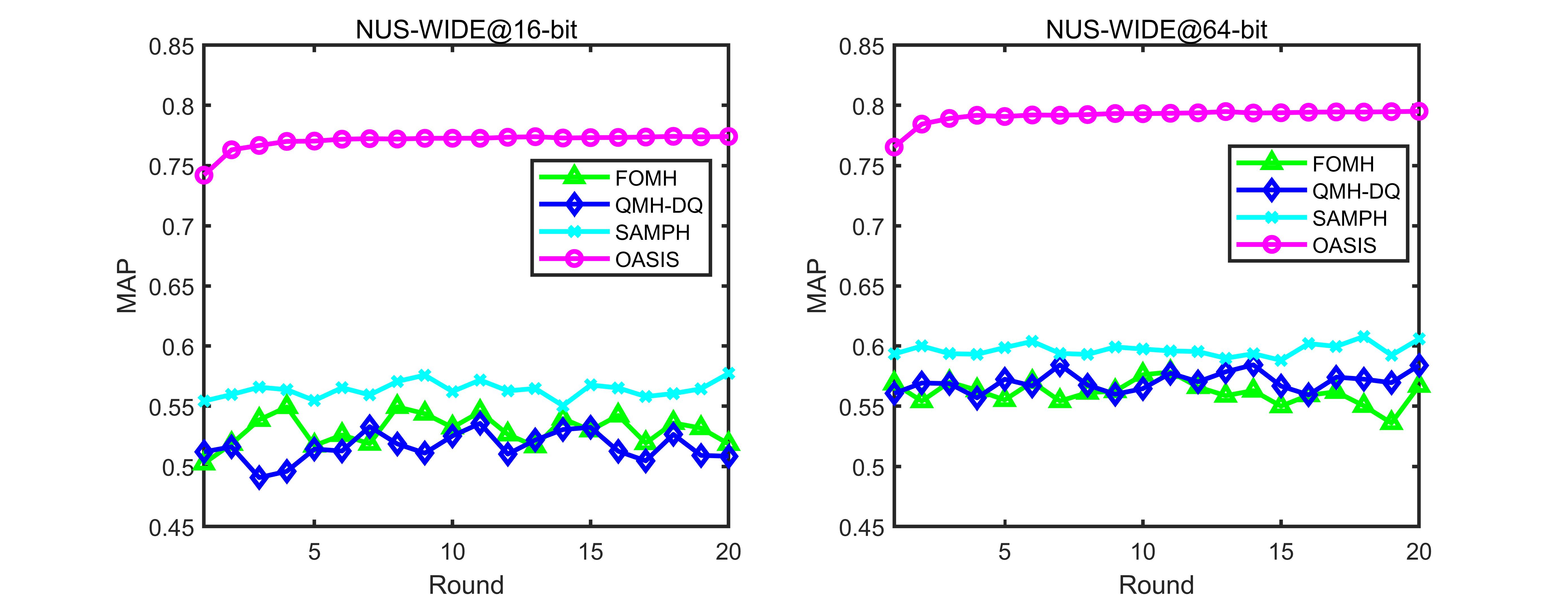}
\caption{ MAP results at all rounds on NUS-WIDE. }
\label{NUS_map}
\vspace{-2mm}
\end{figure}

\begin{figure}[t] 
\centering
\includegraphics[width=0.7\columnwidth]{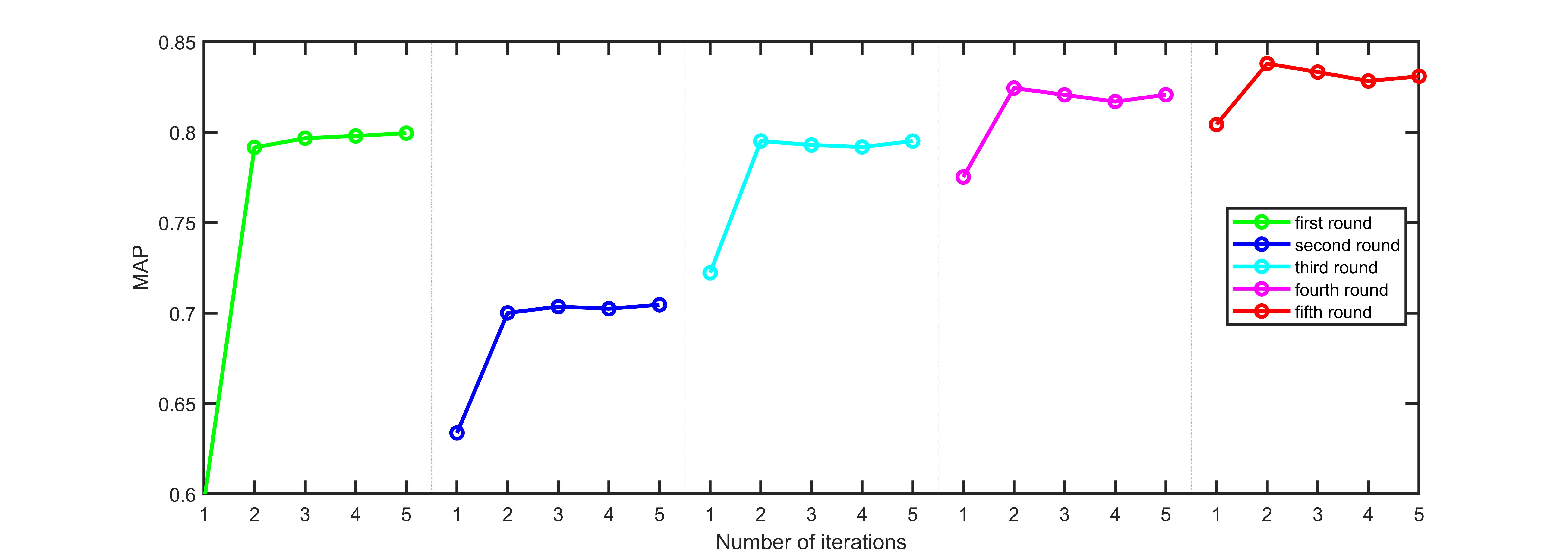}
\caption{Convergence analysis on MIRFlickr.}
\label{convengence}
\vspace{-2mm}
\end{figure}

\begin{figure}[t]
\centering
\includegraphics[width=1\columnwidth]{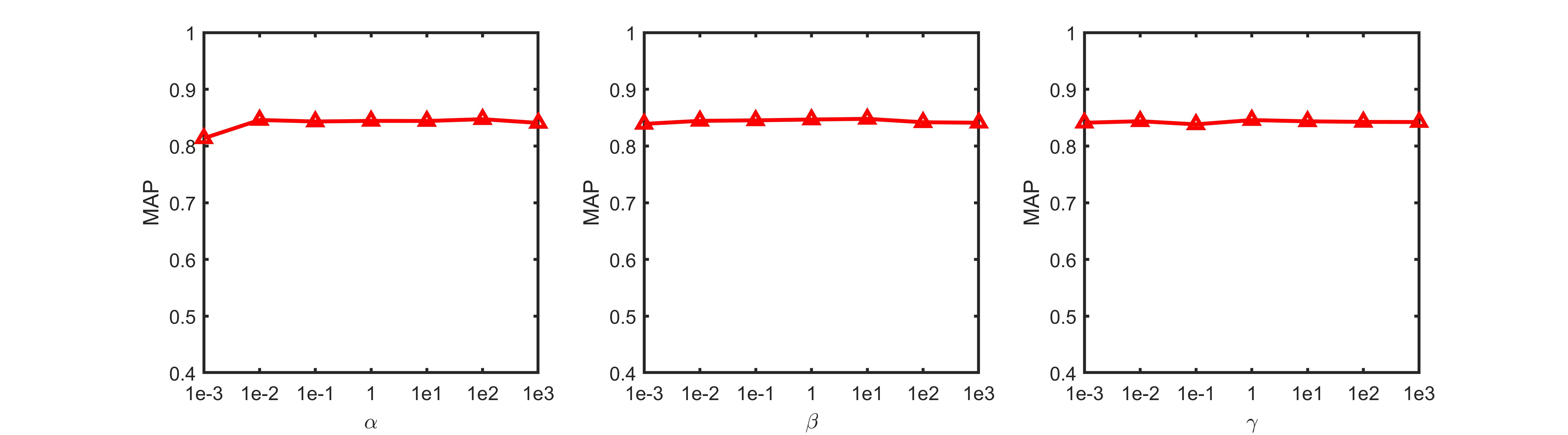}
\caption{Parameter sensitivity of OASIS on MIRFlickr.}
\label{sensitive}
\vspace{-2mm}
\end{figure}

\subsubsection{Baselines.} Three state-of-the-art methods, including FOMH \cite{lu2019flexible},  OMH-DQ \cite{Lu0CNZ19}, and SAPMH \cite{zheng2020adaptive}, are adopted as baselines. To the best of our knowledge, FOMH is the latest and best online multi-modal hashing model. Both OMH-DQ and SAPMH are the most advanced batch-based multi-modal hashing and their hash functions and hash codes are retrained on all accumulated data (i.e., using $\vec{\mathbf{X}}^{(t)}$ and $\widetilde{\mathbf{X}}^{(t)}$) at each round. For all baselines, codes are publicly available and we set their parameters following provided instructions.

As stated in previous section, no existing multi-modal hashing methods try to handle the class incremental problem with streaming data, including our baselines. As the comparisons on the second and third settings still need to be conducted, we had to release the fourth requirement of the problem setting for baselines. That is, we assumed that baselines had already known all categories before training. It is worth noting that our OASIS always meets the four requirements formalized in the Problem Definition Section.

\begin{table*}[]
\centering \footnotesize
\caption{Comparison of training time (seconds) with 16-bit hash code on MIRFlickr.}
\label{time}
\begin{tabular}{llllllllll}
\hline
Methods & round1 & round2 & round3 & round4 & round5 & round6 & round7 & round8 & round9  \\ \hline
FOMH \cite{lu2019flexible}    & 0.3495 & 0.3100 & 0.3100 & 0.3159 & 0.3170 & 0.3239 & 0.3320 & 0.3261 & 0.3200  \\
OMH-DQ \cite{Lu0CNZ19}  & 0.4280 & 0.5692 & 0.6639 & 0.8177 & 1.0162 & 1.1439 & 1.1474 & 1.2908 & 1.6337  \\
SAPMH \cite{zheng2020adaptive}   & 1.6575 & 2.2093 & 2.9318 & 3.3720 & 4.5308 & 5.2791 & 6.4881 & 7.3385 & 10.6951 \\
OASIS   & 1.0948 & 1.0192 & 1.0812 & 1.1807 & 1.2140 & 1.1651 & 1.1831 & 1.1953 & 1.2169  \\ \hline
\end{tabular}
\end{table*}

%\begin{table*}[]
%\centering \footnotesize
%\caption{Comparison of training time (seconds) with 16-bit hash code on MIRFlickr.}
%\label{time}
%\begin{tabular}{clllllllll}
%\hline\toprule[0.5pt]
%Method & chunk 1  & chunk 2  & chunk 3   & chunk 4   & chunk 5   & chunk 6   & chunk 7   & chunk 8   & chunk 9   \\ \hline
%FOMH \cite{lu2019flexible}   & 1.41  & 1.36 & 1.59  & 1.71  & 1.04   & 1.65   & 1.65  & 1.49  & 1.38  \\
%OMH-DQ \cite{Lu0CNZ19} & 5.84  & 5.96  & 6.64  & 7.30 & 8.93  & 8.93  & 10.11 & 11.55 & 13.14 \\
%SAPMH \cite{zheng2020adaptive} & 13.32 & 18.01 & 23.88 & 28.66 & 33.29 & 42.05 & 45.09 & 49.27 & 63.99 \\
%OASIS   & 2.93  & 3.25  & 3.10  & 2.88   & 2.68   & 2.68   & 2.87   & 2.46   & 2.87   \\ \hline\toprule[0.5pt]
%\end{tabular}
%\end{table*}

\subsubsection{Implementation Details}
For OASIS, the parameter settings are: $\alpha =10$, $\beta =1$, $\gamma =1$, $\theta =1$, and $\delta =1 $. We set iteration number as $3$. Our experiments are conducted on a Linux workstation with Intel XEON E5-2650 2.20GHz CPU, 128GB RAM.

\subsection{Results and Comprehensive Analysis}
\subsubsection{MAP Results.}
The MAP values under all experimental settings on two datasets are shown in Table \ref{map_last_round}. We trained all methods with all data chunks and the test set mentioned above is used to retrieve similar samples form database which is consists of all training data that has already arrived. From this table, several observations could be found. 1) The proposed OASIS could always obtain the best performance, demonstrating its effectiveness. 2) In class incremental scenarios, i.e., the second and third settings, our method still works well and the performance gains are significant, validating the motivations and corroborating the effectiveness of our model. 3) The third setting is much more challenging than former two and the performances of all methods drop a lot. Furthermore, we displayed MAP results versus every round under the first setting in Figure \ref{MIR_map} and Figure \ref{NUS_map}. From these figures, the similar observation can be found that our method exceeds the state-of-the-art baselines including batch-based multi-modal hashing and online multi-modal hashing methods.

\subsubsection{Training Time Analysis.}
The training time consumptions of all methods with 16-bit hash code on MIRFlickr are shown in Table \ref{time}. Here, the first setting is adopted. Since there are only 15 training samples at the $10$-th round, we only listed the training time of the first 9 rounds. As OMH-DQ and SAPMH are batch-based models, their hash functions and hash codes are retrained on all accumulated data at each round. Thus, we can find that their training time continuously increases. This phenomenon reflects that batch-based methods fail to efficiently handle streaming data and may be impractical when used in large-scale applications. As FOMH and OASIS are online models, we can find that their training time does not increase with data coming and is only linearly related to the size of new data chunk. Although the time consumption of FOMH is slightly lower, considering the excellent performance of our model, OASIS is more practical and it is worthwhile to get much better accuracy at the cost of a little more time.

\subsubsection{Convergence Analysis.}
The results of the convergence experiments under the first setting are shown in Figure \ref{convengence}. In this figure, we reported the MAP results versus number of iterations at the first five rounds on MIRFlickr in the case of 16-bit code length. It can be seen that our model can quickly achieve satisfactory performance after two iterations.

\subsubsection{Parameters Sensitivity Analysis.}
The experimental results versus different parameter values in the case of 16-bit code length are shown in Figure \ref{sensitive}. We only conducted experiments on $\alpha$, $\beta$, and $\gamma$ because $\theta$ and $\delta$ balance the regularization terms and can be empirically set. In addition, when performing sensitivity experiments on one parameter, we fixed other parameters. It can be seen from the experimental results that when $\alpha$ is small, it has a massive impact on the experimental results so that the semantic pairwise similarity plays a vital role in our model. In addition, we can see that the model is not sensitive to the values of $\beta$ and $\gamma$. Although OASIS has five parameters in all, two of them can be  empirically set and two of  them are not  sensitive. In other words, our method could be easily applied in practical scenarios because its parameters can be readily tuned and selected. We finally set  $\alpha =10$, $\beta =1$, $\gamma =1$, $\theta =1$, and $\delta =1$.

\section{Conclusion and Future Work}
In this paper, we propose a novel hashing method for multi-modal online retrieval, termed Online enhAnced SemantIc haShing (OASIS). OASIS invents a semantic-enhanced representation to describe instances and designs a new objective function to fully learn from the rich semantic information. Besides, with the help of the semantic-enhanced representation, OASIS can handle new classes coming with streaming data well. Sufficient experiments have demonstrated that the performance of our model can surpass the start-of-the-art models. In addition, this paper tries to give explicit task definition and hopes to benefit the hashing community.

Class incremental problem in online multi-modal hashing domain is still challenging. In the future, we will further investigate this problem from the following aspects. 1) Trying to better preserve the knowledge from old classes, especially those don't apper for a long time. 2) Trying to transfer knowledge from old classes to new ones so as to better solve the class incremental problem.

%\subsection{Future Work}
%\subsubsection{Forgotten Problem.}
%In online hashing scenarios, when some categories doesn't appear for a long time, the model may fail to preserve the knowledge about this categories and cause troubles if data with these categories comes for query.
%\subsubsection{Transfer Problem.}
%In online hashing with category increment, knowledge about old categories is unable to fit the data with new categories well because the feature spaces of these two kinds of data are distinct. Maybe we can try to transfer the knowledge from the old-categories data to the new-categories data in future works.

\section{Acknowledgments}
This work was supported in part by the National Natural Science Foundation of China under Grant 61872428, 62172256, in part by Shandong Provincial Key Research and Development Program under Grant 2019JZZY010127, in part by Natural Science Foundation of Shandong Province under Grant ZR2019ZD06, ZR2020QF036, and in part by the Major Program of the National Natural Science Foundation of China under Grant 61991411.

%\clearpage
\bibliography{aaai22}

\end{document}